\documentclass[
 ,twocolumn%
 ,secnumarabic%
,amssymb, amsmath,nobibnotes, aps]{revtex4-1}

\usepackage{bm}
\usepackage{color}

\usepackage{comment}
\usepackage{graphicx}% Include figure files
\usepackage{dcolumn}% Align table columns on decimal point
\usepackage{bm}
\usepackage{mathrsfs}  
\usepackage{extarrows}
\usepackage[table]{xcolor}

\def\bal#1\eal{\begin{align}#1\end{align}}
\newcommand{\p}{\partial}

\usepackage{amssymb,amsfonts,amsmath}

\makeatletter 
\def\@cite#1{(#1)}
\makeatother

\def\t{\widetilde}
\def\d{\delta}
\def\x{{\mathbf x}}

\definecolor{LG}{rgb}{0,0.99,0}
\definecolor{LLG}{rgb}{1,1,0}

\begin{document}

%\date{\today}

\newcommand{\tabenv}[1][\linewidth]{\def\@captype{table}}

\title{Ligand-concentration sensitivity of a multi-state receptor}
% Force line breaks with \\

\author{Takashi Okada  }
% \email{E-mail address: takashi.okada@riken.jp}
\affiliation{%
$^1$Theoretical Biology Laboratory, RIKEN, Wako 351-0198, Japan}

\begin{abstract}
%Receptors detect diffusing molecules with great accuracy to adjust appropriately to environmental changes.  In the classic work by Burg and Purcell, a physical limit of sensitivity of receptors was derived, and  the sensitivity was reexamined for receptors  in a more precise setting by Bialek and Setayeshgar, assuming the system is in thermal equilibrium. However, 

%Previously, the physical limitations of receptor sensitivity have been derived, and this finding was reexamined assuming that the system is in thermal equilibrium, which offers a more precise and stable context.

%Receptors detect diffusing molecules with great accuracy to adjust appropriately to environmental changes. 

Biological sensory systems generally operate out of equilibrium, which often leads to their improved performance. Here, we study the sensitivity of ligand concentration for a general receptor model, which is generally in the non-equilibrium stationary state, in the framework of a stochastic diffusion equation. We derived a general formula of the maximum sensitivity. Specifically,  the sensitivity is limited universally by the Berg-Purcell limit [Biophys. J ., 1977], regardless of whether the receptor is in an equilibrium or non-equilibrium state.

%\begin{description}
%\item[PACS numbers]%02.10.Ud, 47.27.ed, 87.10.-e
%\end{description}

\end{abstract}

%\pacs{9999999 Valid PACS appear here}% PACS, the Physics and Astronomy
   
\maketitle

%\section{Introduction}
Signal detection in biological sensory systems operate with great accuracy. %For example, pheromone receptors can detect a single molecule \cite{LZT}. It is remarkable that sensory systems achieve such high accuracy in fluctuating environments.
A major concern regarding biomolecular sensory systems is the fundamental limitation on sensitivity according to the laws of physics.
The seminal work by Berg and Purcell \cite{BP} proved that the sensitivity of receptors detecting diffusing ligands is limited due to fluctuations in diffusional processes. 

Bialek and Setayeshgar \cite{WS1} improved the argument of the Berg-Purcell (BP) limit more precisely by explicitly including ligand-dissociation/binding processes.  %proving that sensitivity is generally bounded by the BP limit. 
  Following their work and in conjunction with experimental progress, the physical limitations of sensitivity have attracted increased attention in the field of biophysics \cite{WS2,Kaizu, Wingreen1,Wingreen2,Wingreen3,Wingreen4,Wingreen5,Wingreen6,Endres1,Endres2,Mora,Levine1,Levine2,Levine3,Govern,Lang, Fancher}.

In this Letter, we study the sensitivity of ligand concentrations for completely general receptor dynamics. In previous studies \cite{WS1,WS2,Endres1,Endres2}, the system of a receptor  was assumed to be in thermal equilibrium, and the essential theoretical tool used for the arguments was the fluctuation-dissipation theorem (FDT) \cite{Kubo}. However, biological systems are generally out of equilibrium, and many sensory systems utilize free energy dissipations to improve their performance \cite{HF}. Here, we do not assume thermal equilibrium and reexamine the physical limitation of sensitivity for general receptor dynamics, which generally admits a non-equilibrium steady state. By explicitly including all relevant noises in the dynamics, we derive a formula of receptor sensitivity for any single-receptor dynamics. Specifically, we find that  {\it any  non-equilibrium receptor dynamics does not improve the sensitivity beyond the BP limit}, which complements the results of the previous studies based on the FDT.

\begin{figure}[t]
  \includegraphics[width=6cm,bb=0 20 300 150]{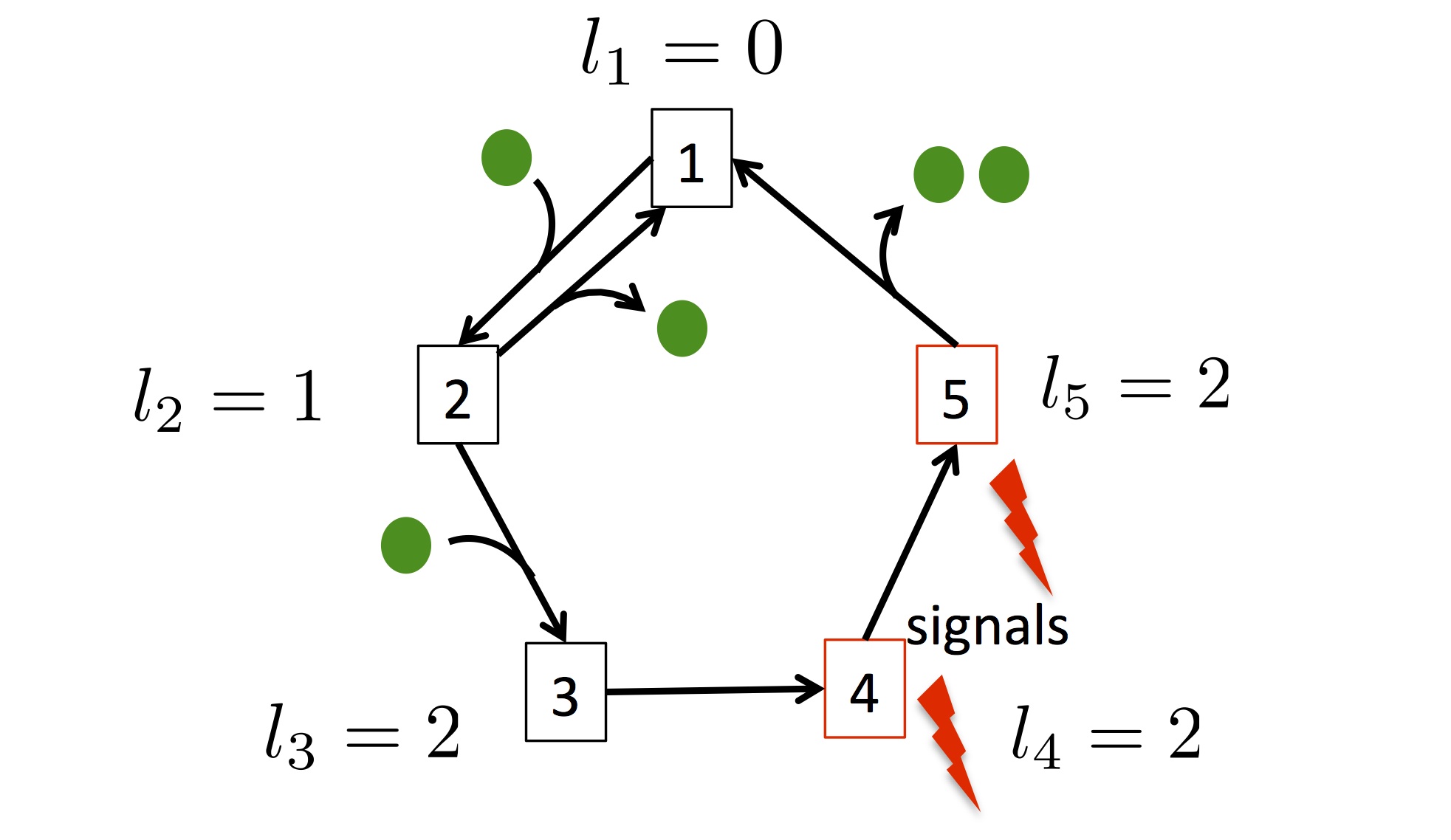}
  \caption{An example of the transition networks of a receptor ($M=5$ and $R=6$). Ligands are represented as green circles and receptor states as square boxes. A subset of the states generates signals indicating ligand concentrations.}
 \label{fig:eg}
\end{figure}

%\section{Receptor with General internal state network}
We consider a receptor with multiple ligand-binding sites and label the receptor states as $m=1,\cdots,  M$ and reactions (transitions among receptor states) as $r=1, \cdots,R$ (Fig. \ref{fig:eg}). We assume that the receptor state jumps from $m=\alpha(r)$ to $m=\beta(r)$ under the $r$-th reaction.
 We introduce the stoichiometric matrix, $\nu$, which is an $M\times R$ matrix whose component is given by 
\bal
\nu_{m,r}=- \delta_{r,\alpha(r)} +\delta_{r,\beta(r)}.\label{nu}
\eal

   %A transition from $m$ to $m'$ with $l_{\beta(r)} -l_{\alpha(r)} >0$ and  $l_{\beta(r)} -l_{\alpha(r)} <0$  corresponds to  a ligand binding and dissociation reaction, respectively.  Transitions with $l_{\beta(r)} -l_{\alpha(r)} =0$ correspond to, for example, pure conformational changes or phosphorylations.

The (deterministic) dynamics of the coupled system of the receptor (at $\x = \bold 0$) and ligand molecules is described by 
\bal
\frac{\p c}{\p t} &= D\nabla^2_{3d} \, c -\delta^{(3d)}(\x)\frac{d}{dt}\sum_m l_m  n_m\label{dcdt_determ},
 \\
\frac{d n_m }{d t}  &= \sum_{r} \nu_{m,r} k_r n_{\alpha(r)},\label{dndt_determ}
\eal
where $n_m(t)$ is the fraction of the $m$-th receptor state ($0\leq n_m(t)\leq 1$), and $k_r$ is the rate constant of the $r$-th reaction. $k_r$ depends on the ligand concentration, $c(\x=0,t)$, at the receptor site if $r$ is a ligand-binding reaction (i.e., $l_{\beta(r)} -l_{\alpha(r)} >0$). $\delta^{(3d)}(\x)$ represents the three-dimensional Dirac delta function.

Suppose that the system is in a steady state specified by $c(\x,t)=\bar c$ and $n_m=\bar n_m$. $\bar n_m$ is determined explicitly as a function of rate constants as $\bar n_m(\{ \bar k_r\})$ by solving \eqref{dndt_determ}, where the ligand-concentration dependence enters implicitly through $\bar k_r =k_r|_{c=\bar c}$.
By linearizing the system around the steady state and including stochastic fluctuations \cite{VK}, we obtain the following Langevin equations: 
\bal
\frac{\p \d c}{\p t} &= D\nabla^2_{3d} \, \d c -\delta^{(3d)}(\x)\frac{d}{dt}\sum_m l_m  \d n_m -{\mathbf \nabla}_{3d} \cdot {\mathbf J}, \label{dcdt} \\
\frac{d \d n_m }{d t}  &= \sum_{r} \nu_{m,r} \biggl(\bar k_r \d n_{\alpha(r)} +  \bar k'_r \bar n_{\alpha(r) }\d c (\x=0,t)\biggr)\nonumber \\
  &+\sum_{r} \nu_{m,r} \xi_r \label{dnmdt}.
\eal
Here,   $\bar k'_r \equiv \p k_r(c)/\p c|_{c=\bar c}$ is nonzero only when the $r$-th reaction is a ligand-binding reaction.
$\xi_r$ represents the noise associated with the $r$-th reaction, satisfying 
\bal
\langle \xi_r(t) \xi_{r'}(t') \rangle =\delta_{r,r'} \bar k_r \bar n_{\alpha(r)} \delta(t-t'),\label{xixi}
\eal
and ${\mathbf J} (t,\x) = (J_x,J_y,J_z)$ is  the diffusional noise, satisfying 
\bal
\langle J_i (t,\x) J_j (t',\x') \rangle  = 2  D \bar c\,  \delta_{i,j}  \delta(t-t') \delta^{(3)}(\x -\x').  \label{JJ}
\eal
The term $-\nabla\cdot J $ in \eqref{dcdt} and \eqref{JJ} can be derived by  regarding the diffusional process as a special type of ``reaction'', where a molecule at a site (in the three dimensional space) is ``produced'' from one located at a neighboring site, and by using van Kampen's size expansion  \cite{Gardiner,ZOS} (see also \cite{Fancher}). 
%We note that $\frac{\p \d c}{\p t}$ in \eqref{dcdt} is affected not only by the diffusional noise but also by reaction noises through $\frac{d}{dt} \d n$.

By applying the Fourier transform to Eqns. \eqref{dcdt} and \eqref{dnmdt}, we obtain
\bal
- i  \omega \sum_{m'}( \delta_{m,m'}-l_{m'} \tau_c)\t{\d n}_{m'}- \sum_r \nu_{m,r}\bar k_r \t{\d n}_{\alpha(r)} \nonumber \\
 = \sum_r \nu_{m,r}[ \bar k'_r \bar n_{\alpha(r)}\t {\mathcal J}+ \t\xi_r],\label{tdn}
\eal
where 
\bal
\tau_c (\omega) &\equiv\frac{1}{\bar c} \int \frac{d^3 k }{(2 \pi )^3} \frac{1}{- i \omega + D {\mathbf k^2} } \approx  \frac{\Lambda}{2\pi^2 D \bar c },\label{tau} \\
 \t{\mathcal J}(\omega) &\equiv  \int \frac{d^3 k }{(2 \pi )^3} \frac{- i  {\mathbf k}\cdot \t {\mathbf J}}{- i \omega + D {\mathbf k^2} }.
\eal
In \eqref{tau}, we have evaluated the integral at low frequency ($\omega \ll D \Lambda^2$) by introducing a UV cutoff, $\Lambda$, corresponding to the inverse of the receptor size as in \cite{WS1}. $\tau_c$ represents the time-scale associated with ligand molecules diffusing around the receptor. 
$ \t{\mathcal J}$ represents the effective diffusional noise ``felt'' by the receptor, satisfying when $\omega \ll D \Lambda^2$, 
\bal
\langle  \t{\mathcal J}(\omega)  \t{\mathcal J}(\omega') \rangle &\approx 2\pi (2 \tau_c \bar c^2 ) \d (\omega+ \omega') \label{JJ0},
\eal
where we used \eqref{JJ} and \eqref{tau}. 

For ligand-concentration sensitivity, a relevant object is the spectral density, $S_{m m'}(\omega)$,  defined as
$
\langle \t{\d n_m}(\omega) \t{\d n}_{m'}(\omega')  \rangle  = 2\pi \,  S_{m m'}(\omega) \, \d (\omega+ \omega').
$
 Although we can straightforwardly compute this from \eqref{tdn}, the analytic computation is difficult for general receptor dynamics. 
For our purpose, we need only the long-term behavior (i.e., $S_{m m'}(\omega=0)$), which can be determined indirectly, as shown below.

In the low-frequency region, by dropping the terms proportional to $\omega$,  \eqref{tdn} can be simplified as 
\bal
- \sum_r \nu_{m,r}\bar k_r \t{\d n}_{\alpha(r)}  \approx  \sum_r \nu_{m,r}( \bar k'_r \bar n_{\alpha(r)}\t {\mathcal J}+ \t\xi_r).\label{tdn_approx}
\eal
In contrast to \eqref{tdn}, it is  no longer possible to invert the left-hand side of \eqref{tdn_approx}, because the coefficient matrix, $\nu_{m,r}\bar k_r$, on the left-hand side is rank-deficient due to the conservation  $\sum_m \frac{d }{dt}n_m=0$. One naive way to avoid this difficulty is to eliminate one  of  the $M$ variables  by using $\t{ \d n}_{m} = -\sum_{m' \neq m} \t{ \d n}_{m'}$ and express \eqref{tdn_approx} in terms of the remaining $M-1$ variables. However, this asymmetric treatment of variables is inconvenient for the derivation of general formulas.

A key step in our approach is to make use of the following relationships satisfied by $ \frac{\partial \bar n_m }{\partial \bar c}$ and  $ \frac{\partial \bar n_m }{\partial \bar k_r}$,  
\bal
- \sum_r \nu_{m,r}\bar k_r  \frac{\partial \bar n_{\alpha(r)} }{\partial \bar c}  &= \sum_r \nu_{m,r} \bar k'_r \bar n_{\alpha(r)}\nonumber,\\
- \sum_r \nu_{m,r}\bar k_r  \frac{\partial \bar n_{\alpha(r)} }{\partial \bar k_{r'}}  &=  \nu_{m,r'}  \bar n_{\alpha(r')} 
,\label{deriv}
\eal 
which can be easily obtained from \eqref{dndt_determ}. 
The comparison of the coefficients in \eqref{tdn_approx} and  \eqref{deriv} implies that \eqref{tdn_approx} can be expressed as  
\bal
\t{\d n}_{m} \approx {  \frac{\partial \bar n_{m}}{\p \bar c}}  \t{\mathcal J}+ \sum_r  \frac{1}{\bar n_{\alpha(r)}}\frac{\partial \bar n_m }{\partial \bar k_{r}} {\tilde \xi}_r .\label{dn_deriv}
\eal 
See the Appendix for a more rigorous derivation of \eqref{dn_deriv}. 
The physical meaning of the step from \eqref{tdn_approx} to  \eqref{dn_deriv} is that, {\it the low-frequency fluctuations $\t{\d n}_{m}(\omega \approx 0)$ can be determined from the dependences of the steady state on external parameters, $\bar c$ and $\bar k_r$}. 
We call the derivatives $ \frac{\partial \bar n_{m}}{\p \bar c}$ and  $ \frac{\partial \bar n_{m}}{\p \bar k_r}$  the  {\it susceptibilities} of the steady states to $\bar c$ and $\bar k_r$, respectively.

Finally, from \eqref{xixi}, \eqref{JJ0}, and \eqref{dn_deriv}, we obtain 
\bal
S_{m,m'}(\omega=0)= 2 \tau_c \bar c^2  \frac{\partial \bar n_m }{\partial \bar c}  \frac{\partial \bar n_{m'} }{\partial \bar c} +S^{reac}_{m,m'}\label{S},
\eal
where 
\bal 
S^{reac}_{m,m'}\equiv \sum_r\frac{ \bar k_r }{\bar n_{\alpha(r)}} \frac{\partial \bar n_m }{\partial \bar k_{r}} \frac{\partial \bar n_{m' }}{\partial \bar k_{r}} \label{sreac}
\eal
represents the contribution from the reaction noises, $\xi_r$. 

\begin{comment}
We note that as in \cite{}, $S^{reac}$ is usually  computed by solving 
the continuous Lyapunov equation, 
\bal
 A S^{reac} + S^{reac} A^T +Q=0, \label{Lyapunov}
\eal
where $A$ is a Jacobian matrix defined as $(A)_{m,m'} = \sum_r \nu_{m,r}\bar k_r \delta_{\alpha(r),m'}$, and $Q$ is given by $(Q)_{m,m'} =\sum_r  \nu_{m,r}\nu_{m',r} \bar k_r \bar n_{\alpha(r)}$. $S^{reac}$ is usually  computed by solving \eqref{Lyapunov}; however, for receptor dynamics \eqref{dndt_determ}, $S^{reac}$ can be obtained more easily and straightforwardly by solving $\bar n(\bar k_r)$ from \eqref{dndt_determ} and using \eqref{sreac}. 
\end{comment}

Similar to \cite{WS1,WS2,Wingreen5,Wingreen6}, we assume that the cell ``averages''  the receptor states over a long-term period, $T$, and quantify the sensitivity of ligand concentration, $\Delta c$, based on the signal-to-noise ratio (SNR). Therefore, we analyze the time-averaged fluctuations 
\bal
\delta N_m \equiv \frac{1}{T} \int\, dt \, \delta n_m (t) \label{dN},
\eal
and the variances
\bal
C_{m,m'} \equiv \langle  \delta N_m  \delta N_{m'} \rangle = \frac{1}{T}  S_{mm'}(\omega =0).
\eal
Suppose that a subset of receptor states ({\it active states}), ${\mathcal M}_{a} \subset \{1, \cdots, M \}$, generates signals indicating the ligand concentration. The maximum SNR is then given by  
\bal
SNR =\sum_{m,m' \in {\mathcal M}_{a}} \frac{\p \bar n_m}{\p \bar c}(C^{-1})_{m,m'} \frac{\p \bar n_{m'}}{\p \bar c}( \Delta c)^2\label{max}.
\eal
The maximum sensitivity (or resolution) can be estimated from the point at which the $SNR$ equals one, which leads to 
\bal
\left(\frac{\Delta c}{\bar c}\right)^2  %&=\frac{1}{\bar c^2} \frac{1}{{\sum_{m,m' \in {\mathcal M}_{a}} \frac{\partial \bar n_m }{\partial \bar c} C^{-1}_{m,m'} \frac{\partial \bar n_{m'} }{\partial \bar c}} }\nonumber \\
& = \frac{1}{T} \frac{1}{\bar c^2}\frac{1}{{\sum_{m,m' \in {\mathcal M}_{a}} \frac{\partial \bar n_m }{\partial \bar c} S^{-1}_{m,m'} \frac{\partial \bar n_{m'} }{\partial \bar c}} }.
 \label{snr}
\eal
By plugging \eqref{S} into \eqref{snr} with some matrix manipulation, the maximum sensitivity becomes 
\bal
\left(\frac{\Delta c}{\bar c}\right)^2 = \frac{2 \tau_c }{ T } + \frac{1}{T \bar c^2}\frac{1}{\sum_{m,m' \in {\mathcal M}_{a}}\frac{\partial \bar n_m }{\partial \bar c}   (S^{reac})^{-1}_{m,m'} \frac{\partial \bar n_{m'} }{\partial \bar c} }.\label{snr_result}
\eal 
The first term is the same as the BP limit, and the receptor kinetics enters into the second term, which is positive-definite, because $S^{reac}$ is a covariance matrix. 
Therefore, we have proven that the sensitivity is bounded by the BP limit, regardless of whether the receptor dynamics is in an equilibrium state or a non-equilibrium state.

If, as is usually assumed, all ligand-binding rates are proportional to $\bar c$, the second term  in \eqref{snr_result} can be written as 
\bal
 \frac{1}{T}\frac{1}{\sum_{r,r' \in l.b.}\sum_{m,m' \in {\mathcal M}_{a}} \bar{k}_r \frac{\partial \bar n_m }{\partial  \bar k_r}  (S^{reac})^{-1}_{m,m'} \bar {k}_{r'} \frac{\partial \bar n_{m'} }{\partial  \bar k_{r'}}  },\label{snr_result_linear}
\eal 
where the summation of reactions, $r,r'$, runs over all ligand-binding reactions (l.b.). By utilizing a  technique developed  in \cite{Mochizuki_main, monomolecular_main,OM_main}, the denominator in \eqref{snr_result_linear}  can be  determined from the state-transition network of the receptor dynamics and expressed as a rational function of rate constants, $\bar k_r$ (see Appendix for details). Such an explicit formula for arbitrary single-receptor dynamics does not exist in the literature. This enables us to evaluate the sensitivity systematically, even for receptors with complex dynamics. 

As an illustration, we first examine a simple receptor model studied by Bialek and Setayeshgar in \cite{WS1}. In this model, the receptor has two  states: a ligand-unbound ($m=1$) and  -bound state ($m=2$). The receptor dynamics is described by  
\bal
\frac{d}{dt}\begin{pmatrix}
n_1\\
n_2
\end{pmatrix}=\begin{pmatrix}
-k_1 & k_{-1}\\
k_{1} & - k_{-1}
\end{pmatrix}\begin{pmatrix}
n_1\\
n_2
\end{pmatrix},\label{Bialekmodel}
\eal 
with $k_1 =k'_1 c$. %The steady state is given by $\bar n_2= \frac{k_{-1}}{\bar k_1+k_{-1}}$.
We assume that the cell ``estimates'' the ligand concentration from $n_2$ (i.e., ${\mathcal M}_{a}=\{2\}$). Note that the resulting sensitivity is the same for ${\mathcal M}_{a}=\{1\}$, because $\delta n_1 =-\delta n_2$. The maximum sensitivity  \eqref{snr_result} becomes
\bal
\left(\frac{\Delta c}{\bar c}\right)^2 = \frac{2 \tau_c }{ T} + \frac{1}{T}\frac{2(\bar k_1+k_{-1})} { \bar k_1 k_{-1} },\label{simplest}
\eal
which agrees exactly with the result derived from the FDT in \cite{WS1}. 
We note that, although the approach based on the FDT gives only the sum of the two terms in \eqref{simplest}, our method determines them separately, which makes clear the physical origins of these two terms: the contribution  from the effective diffusional noise, $\mathcal J$, and  from the reaction noises, $\xi_r$, respectively.

\begin{figure}[t]
  \includegraphics[width=7cm,bb=0 20 300 160]{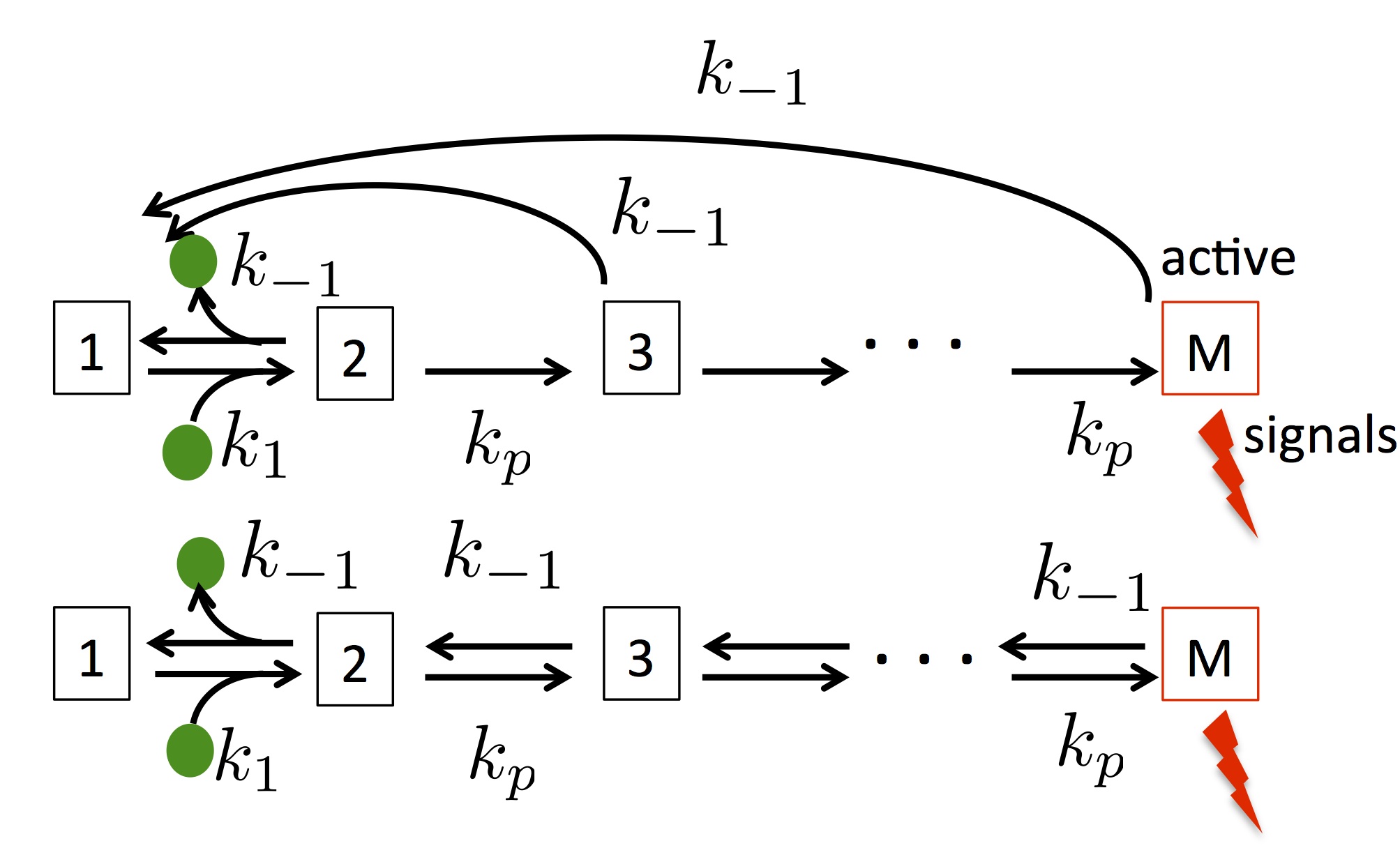}
  \caption{Receptor networks of the kinetic proofreading and reversible models. In both models, we assume that only the $M$-th state is active and generates signals.  }
 \label{fig:chain}
\end{figure}

For more nontrivial and biologically relevant receptor dynamics, we consider a kinetic proofreading model \cite{Mc}  and compare this model with the reversible-reaction analogue (Fig. \ref{fig:chain}). The kinetic proofreading model was originally proposed to explain the ability of T-cell receptors to discriminate foreign antigens from self-antigens based on relatively small differences in ligand affinities. Similar to the kinetic proofreading model of DNA synthesis \cite{HF}, this model utilizes multiple irreversible steps, resulting in large differences in the production of active states depending on affinity.  We remark that we here examine the sensitivity to a single ligand concentration. For a receptor model interacting with spurious ligands, see \cite{Mora}.

In the kinetic proofreading model, the bare receptor binds with a ligand molecule (with rate $k_1 =k' c$), and the ligand-bound state is then phosphorylated up to $M-2$ times (with rate $k_p$ for each modification). The phosphorylated states revert to the unbound state with transition rate $k_{-1}$.
By contrast, the reversible model consists of a ligand-binding reaction (with rate $k_1 =k' c$), $M-2$ forward reactions (with rate $k_p$), and $M-1$ backward reactions (with rate $k_{-1}$).  

  %As  in [McK], in order to reduce parameters, we assume that $\bar c = \frac{k_p}{k'}$, and the ligand dissociation rate is the same as the dissociation rate $k_{-1}$. 

%We assume that $\bar c = \frac{k_p}{k'}$ and the ligand-dissociation rate is the same as the dissociation rate $k_{-1}$. This choices do not change the qualitative behavior of the model. 

We assume that only the final state is active and sends signals indicating ligand concentrations (i.e., ${\mathcal M}_{a} =\{ M\}$). Introducing the dimensionless parameters $\kappa_1,\kappa_{-1}$ as  
\bal
k_1 = \kappa_1 k_p, \ \ k_{-1} = \kappa_{-1} k_p,
\eal
 we can express the maximum sensitivity, \eqref{snr_result}, in the following form:
\bal
\left(\frac{\Delta c}{\bar c}\right)^2 &= \frac{2 \tau_c }{ T } +\frac{  F_M ({\kappa_1,\kappa_{-1}}) }{k_p T},  \label{bothmodel}
\eal
where $  F_{M}$ is a dimensionless factor that depends on $\kappa_1,\kappa_{-1}$ (see the Appendix for the explicit expression of $  F_{M}$). 

%In the proofreading model,  $  F_{M}$ is
%\bal
 % F_{M}=\frac{2(1+\alpha) \{  (1+\alpha)^{M } - M \alpha -1 \}}{ \alpha^3}  \label{prekin},
%\eal
%and, in the reversible model, 
%\bal
 % F_{M}=\frac{2 (\sum_{i=0}^{M-1}\alpha^{i} )(\sum_{i=0}^{2M-4}b_i \alpha^i )}{ \alpha^{M-3}}.
%\label{prerev}
%\eal
%The coefficients $b_i$ are $b_i = \frac{1}{2}(i+1)(i+2)$ for $i\leq M-2$ and $b_i =b_{2M-4-i}$ for $i> M-2$. Note that, for $M=2$, both of \eqref{prekin} and \eqref{prerev} reproduces  \eqref{simplest}.

Before presenting the numerical results, we estimate the two terms in \eqref{bothmodel} for acceptably accurate sensing.  Thus far, we have considered a single receptor. When a cell has many independent receptors, the sensing accuracy of the entire cell is estimated by dividing \eqref{bothmodel} by the total number of  receptors expressed on the cell surface, which we assume to be $\sim 10^4$.  
We estimate $\tau_c=10^{-1} - 10^3 \, {\rm sec}$, (we used $D=10^{-1} -10^1 {\rm \mu  m^2/sec}$, a linear dimension of receptor  $a \equiv \frac{\Lambda}{\pi}\sim 10^{-2} {\rm{\mu m}}$, and  $\bar c = 10^2 - 10^4 /{\rm\mu m^3}$), and the rate constant $k_p =10^{-3}-10^{-1}\, {\rm sec}^{-1}$ (see \cite{para1,para2,para3} for this estimate). 
Using these values, while the first term in \eqref{bothmodel} is acceptably small for the integration times $T\sim \, 10^3\, and {\rm sec}$, the second term can become $\mathcal O(1)$ only if $F_{M}<10^6$. Therefore, in the following, we focus on the receptor-dependent part in \eqref{bothmodel}, $F_M$.

\begin{figure}[t]
  \includegraphics[width=6cm,bb=40 10 205 110]{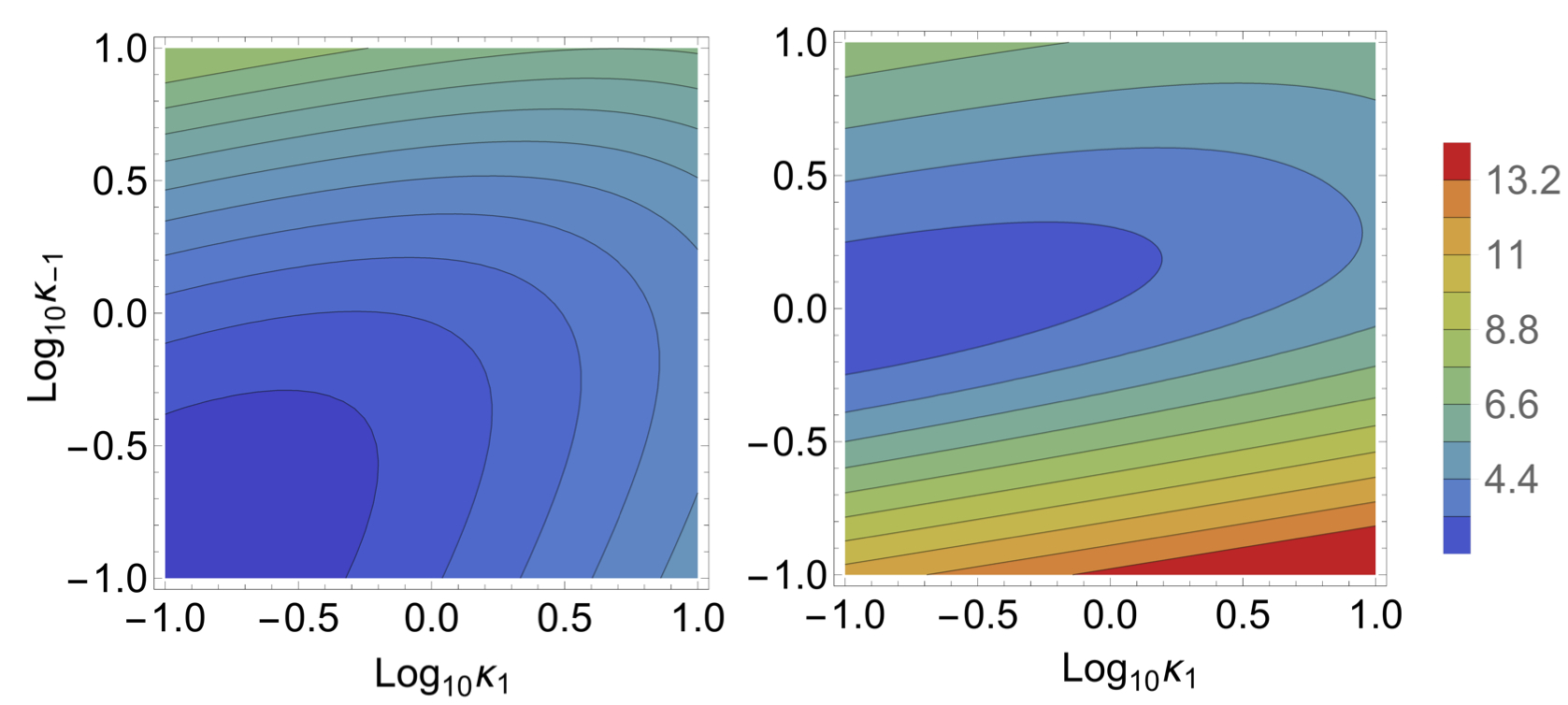}
  \caption{The numerical result of $\log_{10} F_{M}$ for $M=8$ in the kinetic proofreading model (left) and in the reversible model (right). Roughly, $F_{M}<10^6$ is required for accurate estimation of ligand-concentration changes. }
 \label{fig:Na}
\end{figure}

Fig. \ref{fig:Na} shows the numerical results of $  F_{M=8}(\kappa_1,\kappa_{-1})$ in the two models. 
In the region of $\kappa_{-1}>1$ (the upper-half region of Fig. \ref{fig:Na}) corresponding to rapid dissociation, the sensitivities in both models behave in a qualitatively similar way:  $F_M$ is large, except for $\kappa_{-1}\sim 1$, and, as $\kappa_{-1}$ increases, $F_M$ becomes larger (or the sensitivity becomes worse)  rapidly. By contrast, in the region of $\kappa_{-1}<1$, corr esponding to slow dissociation, the behaviors differ qualitatively between the two models. While $  F_{M}$ is large in the reversible models,  $F_{M}$ does not depend significantly upon $\kappa_{-1}$ and remains at a lower level in the kinetic proofreading model. Therefore, when $\kappa_{-1}< 1 $ in  the kinetic proofreading model, an accurate sensing is possible over a wide range of $\kappa_1$ or, equivalently, ligand-concentration, because $\kappa_1 =\frac{k' \bar c}{k_p}$. %We also find that there exists an optimal $\kappa_1$, or equivalently, an  optimal ligand concentration, for each value of $\kappa_{-1}$. 

\begin{figure}[htbp]
  \includegraphics[width=14cm,bb=0 5 380 110]{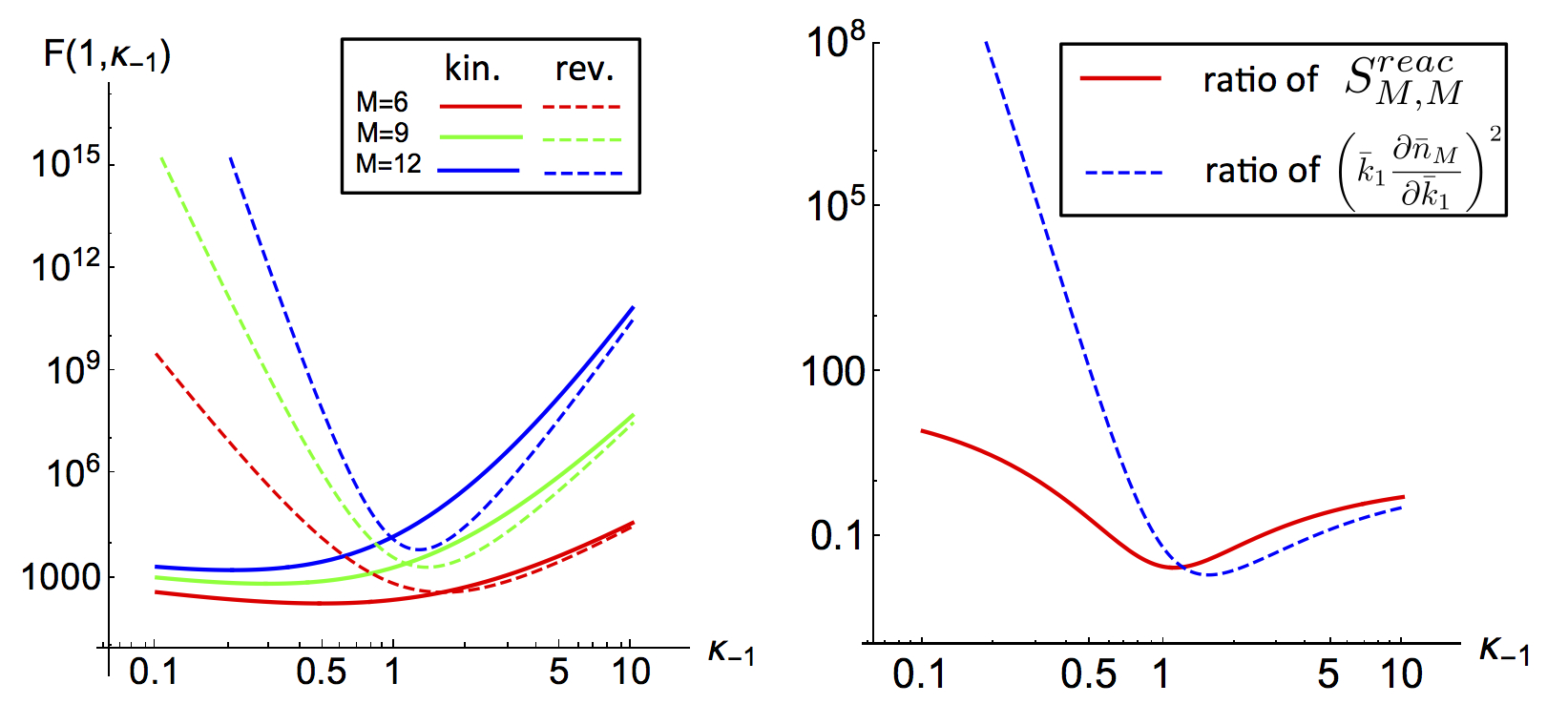}
  \caption{(Left) $F_M(1,\kappa_{-1})$ for $M=6,9,12$ in the kinetic proofreading model (thick lines) and the reversible model (dashed lines). (Right) The thick and dashed lines represent the ratios $X_{kin. \, proof.}/X_{rev.}$ between the two models for $X=S^{reac}_{M,M}$ and $(\bar k_1 \frac{\p \bar n_M}{\p \bar k_{1}})^2 $, respectively.  
  % The thick line represents the ratio between  $S^{reac}_{M,M}$ in the kinetic proofreading model and that in the reversible model, $ S^{reac}_{M,M}|_{kin. proof.}/S^{reac}_{M,M}|_{rev.}$, for $\kappa_{1}=1$ and $M=8$. The dashed line represents the similarly defined ratio for $(\bar k_1 \frac{\p \bar n_M}{\p \bar k_{1}})^2$. A similar result is obtained in other value of $\kappa_1$.%Although $S^{reac}$ behaves qualitatively similarly, the susceptibility differs between the two models. Note that we multiplied  $S^{reac}$ and $\frac{\p \bar n}{\p \bar k_{1}}$  by $\bar k_1$ in order to make them dimensionless. 
   }
 \label{fig:Sreac_dndc}
\end{figure}

 Next, we examine the dependence of $  F_{M}$ on the length of the reaction chains,  $M$ (see Fig. \ref{fig:Sreac_dndc} (Left)). For simplicity of analysis, we set $\kappa_1=1$. From the analytical expression of $F_M$ in the Appendix, 
 we can show that in both models, $F_M$ asymptotically approaches $ F_{M}\sim 2 \kappa_{-1}^{\ M-2}$ when $\kappa_{-1}\gg1$, deteriorating the sensitivity  exponentially as $M$ becomes large. %In particular, when $M\gg 1$, only a tiny  region around $\alpha=1$ is allowable for a precise sensing.
However, when $\kappa_{-1} \ll 1$ and while $F_{M} \sim 2/\kappa_{-1}^{M-3}$ is in the reversible model, which is again exponential in $M$, $F_{M} \sim M(M-1)/\kappa_{-1}$ in the kinetic proofreading model, which depends on $M$ only algebraically. Therefore, when $\kappa_{-1}<1$ and $M$ is large, the sensitivity is much higher in the kinetic proofreading model, compared with the reversible model. Note that in either model, for fixed $\kappa_{-1}$, the sensitivity declines monotonically as $M$ increases.  %Also, the precision of ligand sensing does not deteriorate for a wide region of  $\kappa_{-1}<1$, even in the case that the receptor has many modification steps.

 %The kinetic part of the physical limit in \eqref{snr_result} is determined by $S^{reac}$ and the ligand dependence of the final active state, $\frac{\p \bar n}{\p \bar c}$.  
From where does the discrepancy in performance between the two models originate? The sensitivity  is determined form the ratio between the (squared) susceptibility, $( \bar k_1 \p \bar n_M /\p \bar k_1)^2$, and the fluctuation, $S^{reac}_{M,M}$ (see \eqref{snr_result_linear}). As shown in Fig. \ref{fig:Sreac_dndc} (right), the value of $S^{reac}_{M,M}$ does not differ significantly between the two models. Therefore, the higher accuracy in the kinetic proofreading model essentially derives from its higher susceptibility, which can be understood as follows:  In the reversible model, $\bar n_i/\bar n_{i-1}=\frac{1}{\kappa_{-1}}$ for $i=3,\ldots,M$. Therefore, when $\kappa_{-1}<1$, the dependence of $\bar n_M$ on  $\bar k_1$ diminishes along the long reaction chain, because a large factor, $\frac{1}{\kappa_{-1}}$, is multiplied in each step toward the active state. By contrast, in the kinetic proofreading model, $\bar n_i/\bar n_{i-1}=\frac{1}{1+\kappa_{-1}}$ for $i=3,\ldots,M-1$, which is not large when $\kappa_{-1}<1$. Therefore, the dependence on $\bar k_1$ is maintained along the reaction chain.  

We note that, in the study of T-cell receptors in \cite{Mc}, it is the susceptibility to the dissociation constant, $\p \bar n_M/ \p k_{-1}$, that  leads to T-cell receptor selectivity. However, what we have discussed here is the susceptibility to ligand concentration, $\p \bar n_M/\p \bar k_{1}$,  which is relevant for the sensitivity to ligand concentration.

\begin{comment}
We provide quantitative analysis for the susceptibility, $\frac{\p \bar n}{\p \bar c}$.  In the kinetic proofreading model, we have
\bal
\bar n_M = \frac{k_1 }{(k_1+k_m)(1+\kappa_{-1})^{M-2}}.
\eal
Therefore, when $\kappa_{-1}=\frac{k_m}{k_p}>1$, $\bar n_M$ is highly suppressed for large $M$, as is $\p \bar n_M/\p \bar c$ appearing in the denominator of \eqref{snr_result}. For $\kappa_{-1} <1$, there is no such suppression. In the reversible model, 
\bal
\bar n_M = \frac{k_1}{k_p \kappa_{-1}^{M-1}+k_1 \sum_{i=0}^{M-2} \kappa_{-1}^i}.
\eal
This approaches a ligand-independent constant when $\kappa_{-1}<1$ and becomes small for $\kappa_{-1}>1$. Therefore, the derivative $\p \bar n_M/\p \bar c$ is  small, except for $\kappa_{-1} \sim 1$. In fact, in the case of $k_1 = k_p$,
\bal
\p \bar n_M/\p \bar c \simeq \frac{1}{\sum_{i=-(M-1)/2}^{(M-1)/2} \kappa_{-1}^i},
\eal
which is symmetric under $\kappa_{-1}\rightarrow \kappa_{-1}^{-1}$ and tiny for either $\kappa_{-1}>1$ and $\kappa_{-1}<1$ when $M$ is large.
\end{comment}

In summary, for precise sensing,  the receptor does not allow many intermediate modification steps in the broad range of $\kappa_{-1}$ in the reversible model. However, in the kinetic proofreading model, precise sensing is compatible with many  internal states, as long as $\kappa_{-1}<1$.

In this Letter, we have derived a general formula for sensitivity, \eqref{snr_result}, by explicitly accounting for diffusional and reaction noises and utilizing a similar  method developed in \cite{Mochizuki_main, monomolecular_main,OM_main}. The sensitivity formula \eqref{snr_result} consists of the BP limit and the term  determined from the network topology of receptor dynamics. Our result is novel in that the assumption of thermal equilibrium is not required, and the formula is applicable to any instance of receptor dynamics. 

%Technically, expressing the receptor state fluctuations as  \eqref{dn_deriv} is essential for the proof, which makes transparent that the Burg-Purcell limit arises purely from the diffusion process around the ligand-binding sites without the kinetic of the receptor dynamics.   
The framework of stochastic diffusion equations can serve as the basis for further research into more complex, realistic  ligand-receptor dynamics investigations. For example, a potential generalization is the case where, in addition to the ligand the receptor estimates its concentration, the receptor  is regulated by other (freely diffusing) ligand species. In this case, as shown in Appendix, $S^{reac}$ in  \eqref{snr_result}  is replaced by
\bal
S^{reac}_{m,m'}\rightarrow S^{reac}_{m,m'}+\sum_i 2\tau_i \bar c_i^2 \left(\frac{\p \bar n_m }{\p \bar c_i} \right)^2, \label{sreac_gen}
\eal
where $i$ labels other ligand species with concentration $\bar c_i$ and  diffusion constant $D_i$, and $\tau_i\equiv \frac{\Lambda}{2\pi^2 D_i\bar c_i}$. We can also investigate reacting  ligands by replacing \eqref{dcdt_determ} by reaction-diffusion equations. Another biologically relevant and theoretically  challenging extension involves dynamically interacting receptors, for example, through ligand-regulated oligomerizations, as in the epidermal growth factor (EGF) receptors \cite{HH}.  
We hope to report progress in these directions in the near future.

%As a biologically relevant application, we demonstrated that  the precision associated with reaction noises  in the irreversible model is better than that in the analogous reversible model. However,  the physical limit derived  by Berg and Purcell does exist independently of the non-equilibrium features of receptor dynamics. 

This work was partially supported by the CREST, Japan Science and Technology Agency. We also express our appreciation to Michio Hitoshima, Atsushi Mochizuki, Alan.D. Rendall, and Yasushi Sako for their inspiring discussions related to this work.

%One of the advantages  of our method is that the formula \eqref{snr_result} can be applied, beyond the context of thermal equilibrium,  to a single receptor which admits non-equilibrium steady state. We also emphasize that \eqref{dn_deriv} is  practically useful even for a receptor with a number of internal states because, as we proved,  the fluctuations at low frequency are determined by solving the steady state concentrations. 

%We compared that the physical limits of the kinetic proofreading model and the reversible reaction model. We found that possessing  multiple internal states are allowed only in the kinetic proofreading model. The underlying mechanism  is essentially the same as the original argument of the kinetic proofreading model [];  the final active state is more sensitive to external ligand concentration in the kinetic proofreading model than in the reversible model. 

\nocite{*}


\begin{thebibliography}{99}

%\bibitem{LZT}Leinders-Zufall, Trese, et al. "Ultrasensitive pheromone detection by mammalian vomeronasal neurons." Nature 405.6788 (2000): 792-796.



\bibitem{BP}
H. C. Berg, and E. M. Purcell, Biophysical journal 20.2 (1977): 193-219.
\bibitem{WS1}
W. Bialek, and S. Setayeshgar, Proceedings of the National Academy of Sciences of the United States of America 102.29 (2005): 10040-10045.


\bibitem{Kubo}
R. Kubo,  Reports on Progress in Physics 29.1 (1966): 255.

\bibitem{WS2}
W. Bialek, and S. Setayeshgar, physical Review Letters 100.25 (2008): 258101.

\bibitem{Kaizu}
K. Kaizu {\it et al.}, Biophysical journal 106.4 (2014): 976-985.

\bibitem{Wingreen1}
R.G. Endres, and N. S. Wingreen, Proceedings of the National Academy of Sciences 105.41 (2008): 15749-15754.
\bibitem{Wingreen2}
R.G. Endres, and N. S. Wingreen, Physical Review Letters 103.15 (2009): 158101.
\bibitem{Wingreen3}
T. Mora, and N. S. Wingreen,  Physical Review Letters 104.24 (2010): 248101.

\bibitem{Wingreen6}
Skoge, Monica, Yigal Meir, and Ned S. Wingreen, Physical review letters 107.17 (2011): 178101.

\bibitem{Wingreen4}
V Sourjik, and NS Wingreen, Current opinion in cell biology 24.2 (2012): 262-268.

\bibitem{Wingreen5}
M. Skoge, et al. , Physical Review Letters 110.24 (2013): 248102.

\bibitem{Endres1}
G. Aquino, and R. G. Endres, Physical Review E 81.2 (2010): 021909.

\bibitem{Endres2}
G. Aquino, and R. G. Endres,  Physical Review E 82.4 (2010): 041902.

\bibitem{Mora}
T. Mora,  Physical Review Letters 115.3 (2015): 038102.

\bibitem{Levine1}
W. J. Rappel, and H. Levine,  Proceedings of the National Academy of Sciences 105.49 (2008): 19270-19275.

\bibitem{Levine2}
W. J. Rappel, and H. Levine, Physical Review Letters 100.22 (2008): 228101.

\bibitem{Levine3}
B. Hu, W. Chen, W. J. Rappel, and H. Levine,  Physical Review Letters 105.4 (2010): 048104.

\bibitem{Govern}
C. C. Govern, and P. R. ten Wolde.  Physical review letters 113.25 (2014): 258102.
\bibitem{Lang}
A. H. Lang, C.K. Fisher, and T. Mora, Physical review letters 113.14 (2014): 148103.



\bibitem{Fancher}
S. Fancher, and A. Mugler. Physical Review Letters 118.7 (2017): 078101.


\bibitem{HF}
J. J. Hopfield,  Proceedings of the National Academy of Sciences 71.10 (1974): 4135-4139.

\bibitem{VK}
N.G. van Kampen,  Canadian journal of physics 39.4 (1961): 551-567.

\bibitem{ZOS}
De Zarate, Jose M. Ortiz, and Jan V. Sengers, Hydrodynamic fluctuations in fluids and fluid mixtures. Elsevier Science, Amsterdam Netherlands, 2006.



\bibitem{Gardiner}
Gardiner, Crispin W. Stochastic methods. Springer-Verlag, Berlin-Heidelberg-New York-Tokyo, 1985.


\bibitem{Mochizuki_main}
A. Mochizuki, and B.  Fiedler,  Journal of theoretical biology, 367 (2015), 189-202.

\bibitem{monomolecular_main}
B. Fiedler, and A. Mochizuki, Mathematical methods in the applied sciences, 38 (2015): 3381-3600.

\bibitem{OM_main}
T. Okada, and A. Mochizuki, Physical Review Letters 117.4 (2016): 048101.


\bibitem{Mc}
T. W.  Mckeithan,  Proceedings of the national academy of sciences 92.11 (1995): 5042-5046.

\bibitem{para1}
J. D. Stone, A. S. Chervin, and  D. M. Kranz, Immunology 126.2 (2009): 165-176.

\bibitem{para2}M. Hsieh {\it et al.},  BMC systems biology 4.1 (2010): 57.

\bibitem{para3}
H. Shankaran,  H. S. Wiley, and H. Resat, Biophysical journal 90.11 (2006): 3993-4009.



\bibitem{HH}
C. H. Heldin, Cell 80.2 (1995): 213-223.

\end{thebibliography}

\begin{thebibliography}{99}

%\bibitem{LZT}Leinders-Zufall, Trese, et al. "Ultrasensitive pheromone detection by mammalian vomeronasal neurons." Nature 405.6788 (2000): 792-796.



\bibitem{Mochizuki}
A. Mochizuki, and B.  Fiedler,  Journal of theoretical biology, 367 (2015), 189-202.

\bibitem{monomolecular}
B. Fiedler, and A. Mochizuki, Math. Meth. Appl. Sci. 38 (2015), 3381-3600.

\bibitem{OM}
T. Okada and A. Mochizuki, Law of localization in chemical reaction networks, Physical Review Letters 117.4 (2016): 048101.




\end{thebibliography}
\end{document}